\documentclass[12pt]{solarphysics}

\usepackage[hyperref,optionalrh]{spr-sola-addons_edited}

\usepackage{natbib}
\bibpunct{(}{)}{;}{a}{}{,}
\usepackage{txfonts}

\usepackage{booktabs}

\usepackage[leftcaption]{sidecap}
\usepackage[font=small,labelfont=bf]{caption}

\sidecaptionvpos{table}{t}
\usepackage[labelsep=period]{caption}

\usepackage[left=1.8cm,bottom=1.7cm]{geometry}

\usepackage{url}
\usepackage{siunitx}
\usepackage{enumerate}
\usepackage{pifont}
\usepackage{graphicx}        
\usepackage{wrapfig}
\usepackage{color}           

\chardef\us=`\_

\begin{document}

\begin{article}
\begin{opening}

\title{Linear Polarization Features in the Quiet-Sun Photosphere: \\{\it Structure and Dynamics}}

\author[addressref={aff1,aff2},corref,email={sepideh.kianfar@astro.su.se}]{\inits{S.}\fnm{S.}~\lnm{Kianfar}\orcid{0000-0001-7349-8653}}
\author[addressref={aff3,aff4}]{\inits{S.}\fnm{S.}~\lnm{Jafarzadeh}\orcid{0000-0002-7711-5397}}
\author[addressref=aff5]{\fnm{M.~T.}~\lnm{Mirtorabi}\orcid{0000-0001-9277-3366}}
\author[addressref=aff6]{\fnm{T.~L.}~\lnm{Riethm\"{u}ller}\orcid{0000-0001-6317-4380}}

\address[id=aff1]{Faculty of Basic Sciences, Azad University, P.O. Box 14676-86831, Tehran, Iran}
\address[id=aff2]{Institute for Solar Physics, Department of Astronomy, Stockholm University, AlbaNova University Centre, SE-106 91 Stockholm, Sweden}
\address[id=aff3]{Rosseland Centre for Solar Physics, University of Oslo, P.O. Box 1029 Blindern, NO-0315 Oslo, Norway}
\address[id=aff4]{Institute of Theoretical Astrophysics, University of Oslo, P.O. Box 1029 Blindern, NO-0315 Oslo, Norway}
\address[id=aff5]{Department of Physics, Alzahra University, P.O. Box 1993893973, Tehran, Iran}
\address[id=aff6]{Max Planck Institute for Solar System Research, Justus-von-Liebig-Weg 3, DE-37077 G\"{o}ttingen, Germany}

\runningauthor{S. Kianfar et al.}
\runningtitle{Linear Polarization Features in the Quiet-Sun Photosphere}

\begin{abstract}

We present detailed characteristics of linear polarization features (LPFs) in the quiet-Sun photosphere from high resolution observations obtained with {\sc Sunrise}/IMaX. We explore differently treated data with various noise levels in linear polarization signals, from which structure and dynamics of the LPFs are studied. Physical properties of the detected LPFs are also obtained from the results of Stokes inversions. The number of LPFs, as well as their sizes and polarization signals, are found to be strongly dependent on the noise level, and on the spatial resolution. 
While the linear polarization with signal-to-noise ratio $\geq4.5$ covers about 26\% of the entire area in the least noisy data in our study (with a noise level of $1.7\times10^{-4}$ in the unit of Stokes $I$ continuum), the detected (spatially resolved) LPFs cover about 10\% of the area at any given time, with an occurrence rate on the order of $8\times10^{-3}$~s$^{-1}$\,arcsec$^{-2}$.
The LPFs were found to be short lived (in the range of $30-300$~s), relatively small structures (radii of $\approx0.1-1.5$~arcsec), highly inclined, posing hG fields, and move with an average horizontal speed of 1.2~km~s\textsuperscript{-1}. The LPFs were observed (almost) equally on both upflow and downflow regions, with intensity contrast always larger than that of the the average quiet-Sun.

\end{abstract}

\end{opening}

\section{Introduction}
     \label{S-Introduction}

The magnetic field in the solar photosphere is often inferred by measuring different polarization states (\textit{i.e.}, Stokes parameters; \textit{e.g.}, \citealt{Stenflo1971,Wittmann1974,Auer1977}), from observations of magnetically sensitive lines, such as Fe~{\sc i}~5250.2 \AA. The magnetic structures are distributed all over the solar surface with a variety of temporal and spatial scales, and have a wide range of inclination angles (in respect to the surface normal; \citealt{stenflo08,martinez08b,deWijn09,solanki09,Almeida11}). Among which, linear polarization signal (\textit{i.e.}, transverse component of the magnetic field) is ubiquitously found in active regions (\citealt{chae04,kubo07,cheung08,lites08b}), in quiet areas (\citealt{ishikawa09,lites08b}), but also in polar regions (\citealt{tsuneta08b}) and at the solar limb (\citealt{martin88,lites02}).

\citet{Lites2017} reported the orientation of the magnetic fields in the quiet-Sun photosphere to be dominantly horizontal. However, the internetwork magnetic fields (\citealt{livingston71,livingston75,martin88,Lin95,Lin99}) have been diversely interpreted in the literature as mainly horizontal \citep{lites96,suarez07b,ishikawa09,suarez12}, isotropic \citep{Asensio09,bommier09}, or even predominantly vertical \citep{Stenflo2010,stenflo13}. Some of these interpretations (which are based on Stokes inversions) have been shown to be biased by, \textit{e.g.}, noise-affected Stokes parameters in the quiet Sun \citep{Borrero2011,Borrero2012,Jafarzadeh14,Borrero2015}. We note that Stokes inversions return reliable inclination angles when applied to data with clear Stokes $Q$ and $U$ signal. It is also shown that conclusion between horizontal and quasi-isotropic distribution of the internetwork magnetic field is not straightforward, because the measurements are still limited by present telescopes in terms of their, \textit{e.g.}, spatial resolution and polarimetric accuracy, which prevent detection of magnetic properties of small-scale structures (for more information see \citealt{lagg16,martinez16,Danilovic2016}).

Early observations of the internetwork magnetic fields, reported them as short—lived, horizontally inclined structures, typically smaller than $1''$ near the solar disk-center (\citealt{lites96}), extending to a few arcseconds during their presence on the photosphere (\citealt{pontieu02}). These transient horizontal fields are reported to have strengths on the order of hG (\citealt{lites96,meunier98}). The nearly horizontal component of the magnetic fields, observed by by \cite{harvey07}, also shows seething patterns with variant spatial and temporal scales and an average linear polarization signal of about $10^{-3}$ in the unit of Stokes $I$ continuum ($I_{c}$).

During the last decade, properties of the internetwork horizontal fields have been studied to higher degrees of accuracy using high-resolution observations with, \textit{e.g.}, VTT (\citealt{beck09,khomenco03}), \textit{Hinode} (\citealt{lites08a,ishikawa08b,ishikawa10,Jin09}), and {\sc Sunrise}/IMaX (\citealt{danilovic10}). These studies have confirmed the transitory nature of the linear polarization features (LPFs) with $100-200$~G magnetic strength and typical lifetime and size of $1-10$ minutes and a few arcsec, respectively; but also comparable to granular cells in both temporal and spatial scales (\citealt{ishikawa10,ishikawa11}). 

Horizontally inclined fields are found in both convective upflows and downflows, assumed to be compatible with horizontal magnetic flux (\textit{i.e.}, the magnetic flux of the horizontal magnetic field) whose advection to the surface is due to granular processes (\citealt{pontieu02,Centeno07,suarez07b,Danilovic2016}). Horizontal magnetic flux emerges inside a granule and is moved aside, to intergranular lanes, by plasma as the granule evolves (\citealt{stein06,Centeno07,ishikawa08b,gomory10}). The strong horizontal signal of the magnetic field in granules may manifest the entity of low-lying magnetic loops (\citealt{pontieu02,martinez07,gomory10,jafarzadeh17}). They appear as linear polarization patches in the center or above the granules (\citealt{lites08a,martinez09}). The internetwork fields also tend to become more horizontal in the upper layers of the photosphere (\citealt{stenflo12,Danilovic2016}). The horizontally oriented magnetic fields do not interfere in the granular evolution (\citealt{beck09}), but the low-lying loops can rather reach the upper layers of the solar atmosphere, thus they may contribute to transient heating of the chromosphere and/or the corona (\citealt{Schrijver97,lites08b,ishikawa09}).

Previously, the internetwork horizontal flux has been found to occur more frequently compared to the vertical internetwork field (\citealt{suarez07b,lites08b,beck09,suarez12}). However, recently the high spatial-resolution data from the {\sc Sunrise} 1-m balloon-borne solar telescope \citep{solanki10,barthol11} has provided us with the largest LPF's rate of occurrence of $7\times10^{-4}\;$s$^{-1}\;$ acrsec$^{-2}$ \citep{danilovic10}, larger by 1--2 orders of magnitude than previously reported.

In this study we aim at investigating the effect of the noise level on the number of identified LPFs, and hence, on their statistical properties, namely, size, lifetime, linear polarization signal, proper motion, line-of-sight (LOS) velocity, field strength, inclination angle, and temperature, by extending the work of \citet{danilovic10}. We provide a thorough overview of properties of LPFs in the quiet-Sun internetwork from high-resolution spectropolarimetric observations obtained with \textit{Imaging Magnetograph eXperiment} (IMaX; \citealt{martinez11}) on board the {\sc Sunrise}, also form the results of Stokes inversions. We analyze differently treated images of the same data set (\textit{i.e.}, non-reconstructed, spatially-smoothed non-reconstructed, and reconstructed; Section~\ref{S-Observ}) to identify features based on various signal-to-noise ratios (S/N). The characteristics of the detected features are provided in Section \ref{S-Analys}, and the concluding remarks are summarized in Section \ref{conculsion}.

\section{Observations and Data Preparation}
	\label{S-Observ}
	We study a quiet-Sun area of $45''\times45''$ at disk center acquired by {\sc Sunsrise}/IMaX on 2009 June 9 (between 01:32 and 01:58 UT), with sampling and temporal resolutions of $0.0545$~arcsec/pixel and 33~s, respectively. The effective field-of-view (FOV) under study was created after removing the edges of the original $51''\times51''$ images, influenced by the apodization effect.
The data set includes full Stokes ($I$, $Q$, $U$, and $V$) observations sampled in five wavelength positions ($\pm80$ and $\pm40$ m\AA\ in the line, and $+227$ m\AA\ as the continuum) around the magnetically sensitive line Fe~{\sc i} centered at 5250.2 \AA.\

We aim at detecting LPFs (\textit{i.e.}, contiguous pixels with considerable linear polarization signals) in these quiet-Sun data, which largely includes relatively weak polarization signals, particularly in the Stokes $Q$ and $U$ (\citealt{Borrero2011,Danilovic2016}). The less noisy the linear polarization maps become, the larger number of features, or features with larger sizes, are detected (\textit{i.e.}, the signals are uncovered to a larger extent; \citealt{suarez12}). Thus, to investigate the effect of noise in identification of these features, we inspect differently treated data of the same time-series with various levels of noise (see below).

The data were prepared through a set of instrumental corrections for, \textit{e.g.}, temporal intensity fluctuations, interference fringes and dust particles in optical elements (including dark and flat-fielding), along with minimizing jitter-induced residual signal and instrument-caused blue-shifts (\citealt{martinez11}). These produced the so-called non-reconstructed data (NR; level 1), with $1\sigma$ noise levels of $8.3\times10^{-4}~I_{c}$ and $1.1\times10^{-3}~I_{c}$ in Stokes $Q$ and $U$, respectively \citep{Jafarzadeh14}. The noise levels of Stokes $Q$ and $U$ were determined as the standard deviations at the continuum positions of the corresponding Stokes parameter, since no significant polarization signal is expected in their continuum.

Furthermore, this data product were passed through a phase diversity (PD) procedure according to the point-spread function (PSF) of the instrument/telescope which amplifies signal frequencies (\textit{i.e.}, PD-reconstructed data, PDR; level 2; \citealt{martinez11}). The latter procedure resulted in images with a higher spatial resolution (by a factor of two; reaching to $0.15-0.18$~arcsec) compared to the NR data, but also with a larger noise level (by a factor of $\approx3$). The Stokes $Q$ and $U$ of the PDR data have noise levels of $2.6\times10^{-3}~I_{c}$ and $3.6\times10^{-3}~I_{c}$, respectively.

The linear polarization maps (we will refer to these images as transverse magnetograms) are then constructed from individual Stokes $Q$ and $U$ signals (from both NR and PDR data) at each wavelength position $i$ as $\sqrt{Q_{i}^2+U_{i}^2}$. The net linear polarization (LP; \citealt{lites08b,martinez11}) is then formed by averaging the four wavelength positions inside the Fe~{\sc i} 5250.2 \AA\ line, normalized to $I_{c}$ of each corresponding pixel, as below
\begin{equation} 
	LP={\frac{1}{4I_{c}}} \sum\limits_{i=1}^4 \sqrt{Q_{i}^2+U_{i}^2}.
	\label{eq:LP}
\end{equation}

The LP has half of the noise of the linear polarization at any individual wavelength position (\textit{i.e.}, $\sigma_{LPind}/\sqrt{4}$). We note that the noise level of individual wavelength positions ($\sigma_{LPind}$) was determined as the standard deviation of the linear polarisation formed at the continuum position (\textit{i.e.}, $\sigma_{\sqrt{Q_{continuum}^2+U_{continuum}^2}}$). Thus, $1\sigma$ noise levels of $3.4\times10^{-4}~I_{c}$ and $1.1\times10^{-3}~I_{c}$ are obtained for LP constructed from NR and PDR data sets (\textit{i.e.}, LP\textsubscript{NR} and LP\textsubscript{PDR}), respectively.

\begin{figure}[h!]
   \centerline{\hspace*{0.05\textwidth} 
   \includegraphics[width=1.2\textwidth,clip=]{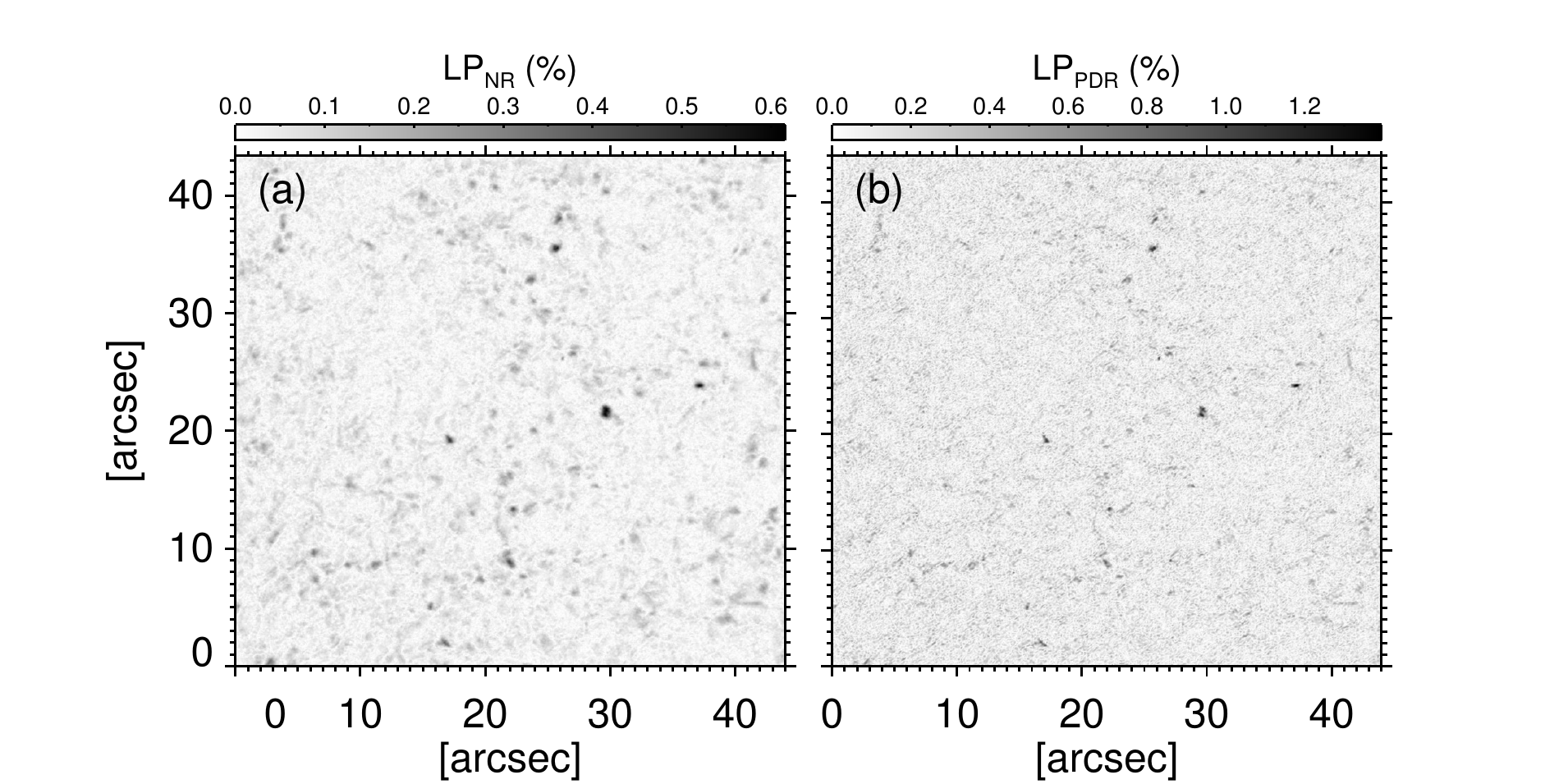}}
     \vspace{0.01\textwidth} 
	\caption{Examples of LP maps (see main text) from the first frame of (a)~non-reconstructed and (b)~phase-diversity reconstructed data. The maps in panels (a) and (b) have $1\sigma$ noise levels of $3.4\times10^{-4}~I_{c}$ and $1.1\times10^{-3}~I_{c}$, respectively. 
        }
   \label{LP_maps}
\end{figure}

Figures~\ref{LP_maps} (a) and (b) display the LP maps of the first frame of NR and PDR data, respectively. The LP\textsubscript{PDR} illustrates sharper and smaller patches compared to that from NR data. Signal amplification of the PD reconstruction has increased the polarization signal by a factor of $2.5-3$ compared to the one in the non-reconstructed map, but also had enhanced the noise level with about the same factor. We find that about the same number of pixels pose signals greater than, or equal to, the noise level in both LP maps in Figure~\ref{LP_maps}. Also, the LP\textsubscript{NR} includes only 14\% more pixels with S/N$\geq4.5$ compared to the LP\textsubscript{PDR}.

To additionally increase the S/N in the linear polarization signals, similar to \cite{Jafarzadeh14}, we spatially smooth the non-reconstructed (SSNR) data by boxcar averaging of individual images (per Stokes parameter and per wavelength position) with a size of 3 pixels (\textit{i.e.}, averaging over 9 pixels without degrading the spatial resolution), prior to forming the LP. This smoothing results in single wavelength-position noise levels of $4.6\times10^{-4}~I_{c}$ and $4.8\times10^{-4}~I_{c}$ in Stokes $Q$ and $U$, respectively, and $1\sigma$ noise level of $1.7\times10^{-4}~I_{c}$, in LP\textsubscript{SSNR}. We note that the noise level of an LP is computed as the standard deviation of the corresponding linear polarization map computed for the continuum position, \textit{i.e.}, standard deviation of $\sqrt{Q_{c}^2+U_{c}^2}$, where $Q_{c}$ and $U_{c}$ are the Stokes $Q$ and $U$ of the associated dataset at the continuum position.

The latter noise level (\textit{i.e.}, from LP\textsubscript{SSNR}) is the smallest in all the differently treated data sets introduced above, that is smaller than those from the NR and PDR data by a factor of $\approx2$ and $\approx6.5$, respectively.

\begin{figure} 
   \centerline{\hspace*{0\textwidth} 
    \includegraphics[width=0.84\textwidth,clip=]{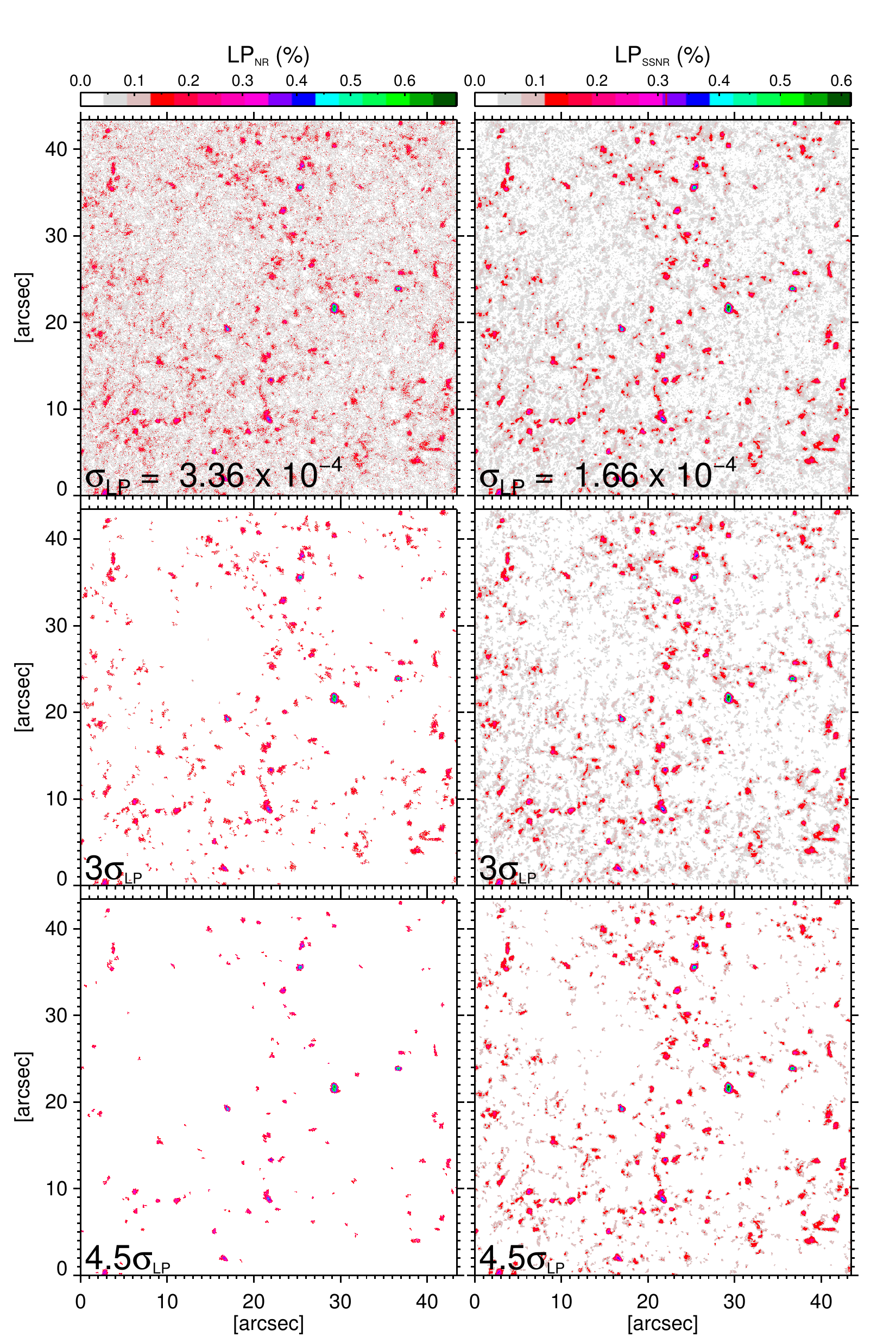}}
     \vspace{0\textwidth} 
	\caption{Examples of LP maps obtained from the first frame of non-reconstructed (\textit{left column}) and spatially smoothed non-reconstructed (\textit{i.e.}, by boxcar averaging of $3\times3$ pixels; \textit{right column}) data from {\sc Sunrise}/IMaX. From top to bottom, linear polarization signals above a certain threshold, labeled in the lower bottom corner of the panels, are plotted.
        }
   \label{features}
\end{figure}

Figure~\ref{features} illustrates the effect of smoothing procedure on the S/N of a single LP map obtained from the first frame of data. Images on the right column show LP\textsubscript{SSNR} maps and the ones on the left represent the LP\textsubscript{NR} ones. The thresholds, as a factor of the corresponding noise levels, above which the LP signals are plotted are indicated on the images (\textit{i.e.}, labels on their lower left corners). As it is visually evident in Figure~\ref{features}, LPFs in smoothed maps are greater in both number and size, particularly in the second and third rows in Figure~\ref{features} (corresponding to minimum S/N of 3 and 4.5, respectively). Pixels satisfying S/N$\geq4.5$ cover about 26\% of the entire LP\textsubscript{SSNR} map, which is larger by a factor of two compared to its non-smoothed counterpart (\textit{i.e.}, the LP\textsubscript{NR} map). We note that a lower signal threshold results in a larger area covered by the linear polarization signal. However, the polarization signal is influenced, to a larger extent, by noise when pixels with smaller S/N are included.

Table~\ref{table:noiselevels} summarizes the various noise levels of the differently treated data, discussed above. For comparison, we will search for LPFs in all the three treated datasets of LP\textsubscript{PDR}, LP\textsubscript{NR}, and LP\textsubscript{SSNR}. However, the LP\textsubscript{SSNR} (with the lowest noise level in our sample) is considered as the primary dataset, hence it provides the main results of this study.

\vspace{3mm}
\begin{SCtable}[][h!]
\small
\caption{Summary of $1\sigma$ noise levels of linear polarization signals in differently treated datasets from {\sc Sunrise}/IMaX (PDR: Phase-diversity reconstructed data; NR: Non-reconstructed data; SSNR: Spatially smoothed non-reconstructed data).} 
\label{table:noiselevels}
\begin{tabular}{l c c c}      
\hline
Parameter & PDR & NR & SSNR\\
\specialrule{.1em}{.05em}{.05em}
\\ [-2.0ex]
   $Q/I_{c}$ & $2.6\times10^{-3}$ & $8.3\times10^{-4}$ & $4.6\times10^{-4}$\\   
   $U/I_{c}$ & $3.6\times10^{-3}$ & $1.1\times10^{-3}$ & $4.8\times10^{-4}$\\
   LP & $1.1\times10^{-3}$ & $3.4\times10^{-4}$ & $1.7\times10^{-4}$\\
\hline
\end{tabular}
\end{SCtable}

\section{Analysis and Results}
	\label{S-Analys}

We aim to search for LPFs in the LP maps from the three differently treated datasets (introduced in Section~\ref{S-Observ}), and for the entire observing duration (consisting of 58 frames). The LPFs are defined as contiguous areas of minimum 10 pixels (\textit{i.e.}, the spatially resolved features) with considerable LP signals (\textit{i.e.}, greater than a specific noise level; \citealt{danilovic10}). We examine a variety of S/N, ranging from 1 to 6.5, as the signal threshold, of which S/N=4.5 is considered as the primary signal criterion since it results in a bigger number of features with precise boundaries.

The LPFs are found to be highly dynamic. During the course of their lifetimes, they face, at least, two of the following phenomena: (a)~emergence, (b)~submergence, (c)~merging, and (d)~splitting. Emergence and submergence of the patches are simply defined as rising and dropping of the signal above and below the noise level, respectively. It is noted that the features that are already emerged in the first frame are considered as new-born features. In addition, the ones that only appear in the last frame, are assumed to have a lifetime of only one frame. We also treat the product of both merging and splitting interactions as new-born features and consider parent features to be dead.

Properties of the identified LPFs (see below for the detection criteria), namely, size as well as average and maximum LP signals are determined in individual frames, and the lifetime and horizontal velocity of the features are calculated in the time-series of LP images. We further obtain their other physical properties, such as magnetic-field strength, field inclination angle, LOS velocity, and temperature from the results of an Stokes inversion code. We provide distribution of the parameters and compare them by means of scatter plots between pairs of quantities.

All the above procedures are applied on the LP\textsubscript{PDR}, LP\textsubscript{NR}, and LP\textsubscript{SSNR} maps. Thus, we inspect the effect of noise level (or S/N) on the number of detected features, and consequently, on their determined properties.

\subsection{Detection and Tracking Procedures}
    \label{Sb:detection}

To detect LPFs in the differently treated LP maps, we employ an algorithm developed by \citet{Jafarzadeh2015}. In detection procedure, the LP signal in a given LP map is cut at a certain signal threshold, which is defined as a factor of noise level (see Figure~\ref{features}). Then, using the IDL routine \verb!blob_analyzer.pro! (\textit{i.e.}, an algorithm for analyzing the regions of interest or the so-called "blobs" in an image), all contiguous remaining (non zero) pixels are identified and given an identity. Their properties, such as area (\textit{i.e.}, the number of pixels that each LPF includes), LP signal, and coordinates of the pixels within individual features are stored. In the detection process, the features are assumed to have circular shape. Therefore, the LPFs smaller than a size threshold of 10 pixels, that leads to a diameter of $\approx0.2$~arcsec (which is approximately equal to the resolution limit of the telescope), are excluded from the sample. Thus, the detected features are not noise originated, and are spatially resolved.

We examine several signal thresholds on the LP maps (of differently treated), ranging from 1$\sigma$ to 6.5$\sigma$, to find a proper S/N which results in the greatest number of detected features that have the most precise boundaries and are not noise originated. An S/N~$\geq4.5$ is found to be the best choice. For comparison, we also present our analysis results for an S/N~$\geq3$, but for simplicity, only for the LP\textsubscript{NR}. The latter will show how the number, and properties, of detected features (of the same data set; with the same noise level) depends on the threshold above which the LPFs are defined.

Furthermore, a new set of image sequences including only LPFs are created (\textit{i.e.}, areas outside the identified features are set to zero). These series of images are then used to track the LPFs in time. We use the same tracking algorithm as described by \citet{Jafarzadeh13}. In this method, the code locates each LPF by determining the center-of-gravity of its intensity as the position of the feature. Considering the location and size of the feature, a small area around each LPF is searched in consecutive frames. We note that only features above a certain signal threshold are included. Therefore, the ``absent allowance'' (where a feature disappears in a few frames during the course of its lifetime, as a result of signal variation with time), which was considered for magnetic bright points in \citet{Jafarzadeh13}, is not performed here. Thus the tracking procedure returns lifetime and horizontal velocity of the LPFs, in addition to the LP signal, size and location of the features which were already stored in the detection phase.

Table~\ref{table:LPFnumbers} summarizes the average number of detected LPFs per frame (\textit{i.e.}, an FOV of $45''\times45''$), the mean fraction of area covered by all the detected LPFs in each frame, and the number of individual LPFs tracked in the entire time-series of LP images, consisting of 58 frames (\textit{i.e.}, when each LPF is counted once during the course of its lifetime). These statistics are listed for the differently treated data sets. The signal thresholds, above which the LPFs are identified, are also specified.

\vspace{3mm}
\begin{table*}[!h]
\small
\caption{Statistics of the number of detected linear polarization features (LPFs) in differently treated data sets from {\sc Sunrise}/IMaX, containing signals above a given threshold.}
\label{table:LPFnumbers}                  
\begin{tabular}{l c c c c c}   
\hline
Data$^*$ & Signal & Number of & Area coverage & Number of & Rate of occurrence\\
set & threshold & LPFs per frame & per frame & individual LPFs$^{**}$& (s$^{-1}$\,arcsec$^{-2}$)\\
\specialrule{.1em}{.05em}{.05em}
\\ [-2.0ex]
   PDR & 4.5$\sigma$\textsubscript{LP\textsubscript{PDR}} & 544 & 0.3\% & 1942 & $6.1\times10^{-4}$\\   
NR & 3$\sigma$\textsubscript{LP\textsubscript{NR}} & 3479 & 5.0\% & 14524 & $4.6\times10^{-3}$\\
   NR & 4.5$\sigma$\textsubscript{LP\textsubscript{NR}} & 535 & 1.1\% & 4092 & $1.3\times10^{-3}$\\
   SSNR & 4.5$\sigma$\textsubscript{LP\textsubscript{SSNR}} & 2073 & 10.3\% & 25099 & $7.9\times10^{-3}$\\
\hline
\multicolumn{6}{l}{$^{~*}$ PDR: Phase-diversity reconstructed; NR: Non-reconstructed; SSNR: Spatially smoothed}\\
\multicolumn{6}{l}{\hspace{2.5mm} Non-reconstructed.}\\
\multicolumn{6}{l}{$^{**}$ When each LPF was counted once during its lifetime.}
\end{tabular}
\end{table*}

The number of individual LPFs found in the SSNR maps are, on average, 4~times larger than those detected in the NR images, when the same signal threshold of $4.5~\sigma_{LP}$ (of the corresponding data) is applied. Note that the total fraction of area covered by the detected LPFs in the LP\textsubscript{SSNR} are larger by an order of magnitude than those found in the LP\textsubscript{NR}.
The number of LPFs in the LP\textsubscript{NR} are comparable to those found in the noisier (but with a higher spatial resolution) LP\textsubscript{PDR} maps. The fraction of FOV covered by all the LPFs in one frame are, however, smaller, by a factor of 3, in the latter data set. These may imply the effect of both spatial resolution and noise level on the number of identified features. Although in NR data several LPFs are detected as one feature, while they are identified individually resolved and separated in the LP\textsubscript{PDR} images in case of applying the same signal threshold, \textit{i.e.}, 4.5$\sigma$, the LP\textsubscript{NR} images have a larger S/N, which results in a larger amount of identified signal (and hence, a larger number of individual features) compared to those from the LP\textsubscript{PDR} maps (see Table~\ref{table:LPFnumbers}).

A comparison between the number of LPFs identified in the LP\textsubscript{NR} images with minimum 3$\sigma$\textsubscript{LP\textsubscript{NR}} and 4.5$\sigma$\textsubscript{LP\textsubscript{NR}} noise levels clarifies that the lower the signal threshold becomes, the larger number of features (and the bigger areal coverage) is obtained. Although the number of detected LPFs in the LP\textsubscript{SSNR} would similarly increase if we would use a lower signal threshold of, \textit{e.g.}, 3$\sigma$\textsubscript{LP\textsubscript{SSNR}}, the final results could be biased by the effect of noise. Clearly, the number of detected LPFs is a trade-off between the necessity of high S/N of the data, as well as the threshold with which the features are defined. Thus, we rather choose a conservative signal threshold of 4.5$\sigma$ in our analysis. We note that our size threshold of 10 pixels has limited the number of detected LPFs, hence, a lower limit of their rates of occurrence are obtained. This has, however, resulted in studying the spatially resolved features.

\subsubsection{Intensity Distribution}
   \label{Sb:F-intensity}

\begin{figure}
   \centerline{\hspace*{0.01\textwidth} 
                          \includegraphics[width=1.0\textwidth,clip=]{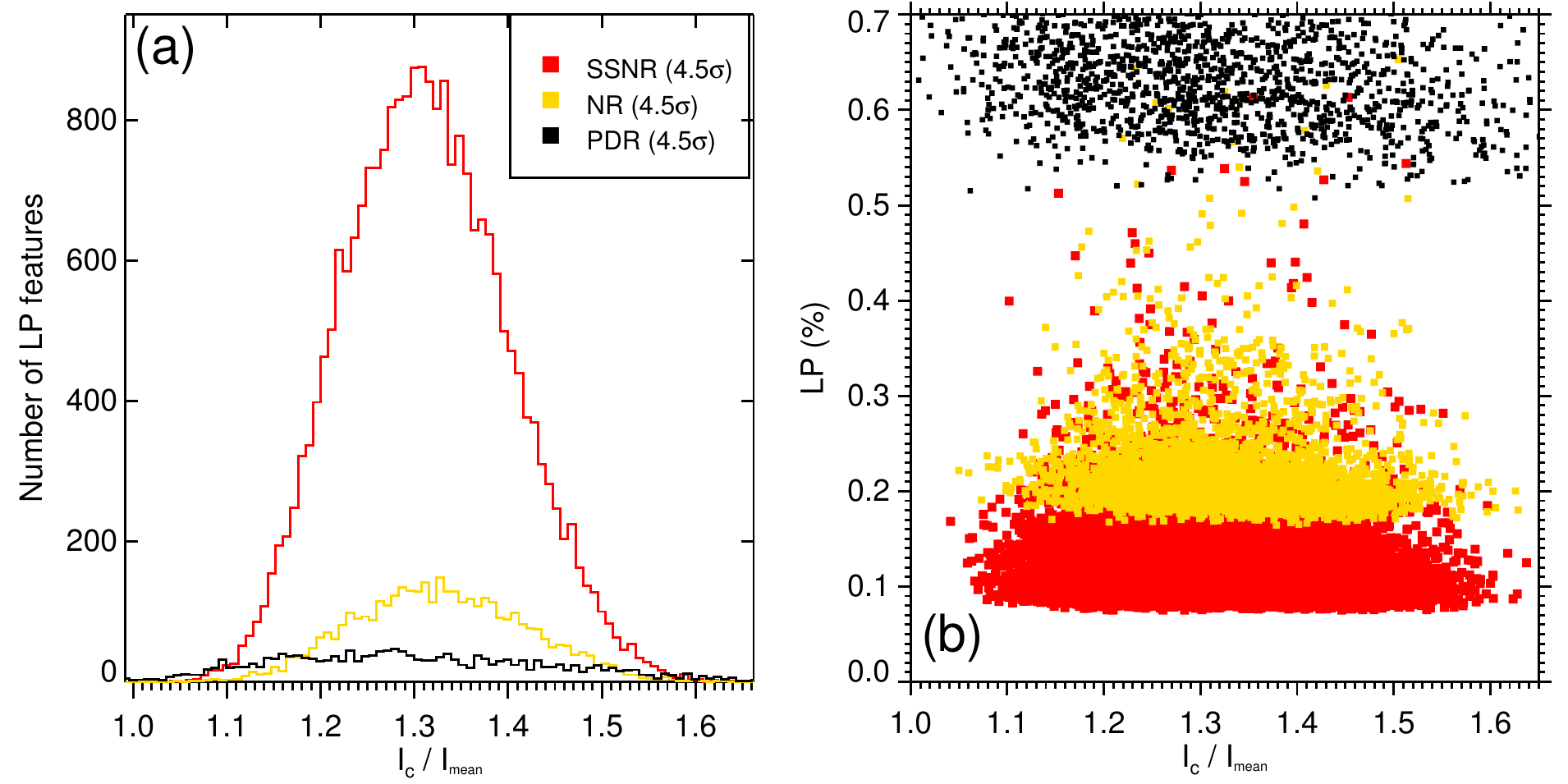}
              }
     \vspace{0.01\textwidth} 
	
	\caption{Distribution of intensity contrast of the detected LPFs (a), and mean linear polarization of the LPFs as a function of their contract (b). The results from three differently treated data sets (PDR: Phase-diversity reconstructed data; NR: Non-reconstructed data; SSNR: Spatially smoothed non-reconstructed data) are plotted.}
   \label{histIc}
\end{figure}

Distributions of the mean intensity contrast of all detected LPFs from Stokes $I$ continuum (normalized to the average quiet-Sun) from SSNR, NR, and PDR images are illustrated in Figure~\ref{histIc}(a). These histograms show that the LPFs are all brighter than the average quiet-Sun ($I_{c}=1$). The relationships between these intensity contrast values and their mean LP signals are also plotted in Figure~\ref{histIc}(b). The latter shows no relationship between the brightness and LP signal of the features under study.

\subsubsection{Physical and Dynamical Properties}
   \label{Sbb:F-char}

\begin{figure}
   \centerline{\hspace*{0.01\textwidth} 
                          \includegraphics[width=0.98\textwidth,clip=]{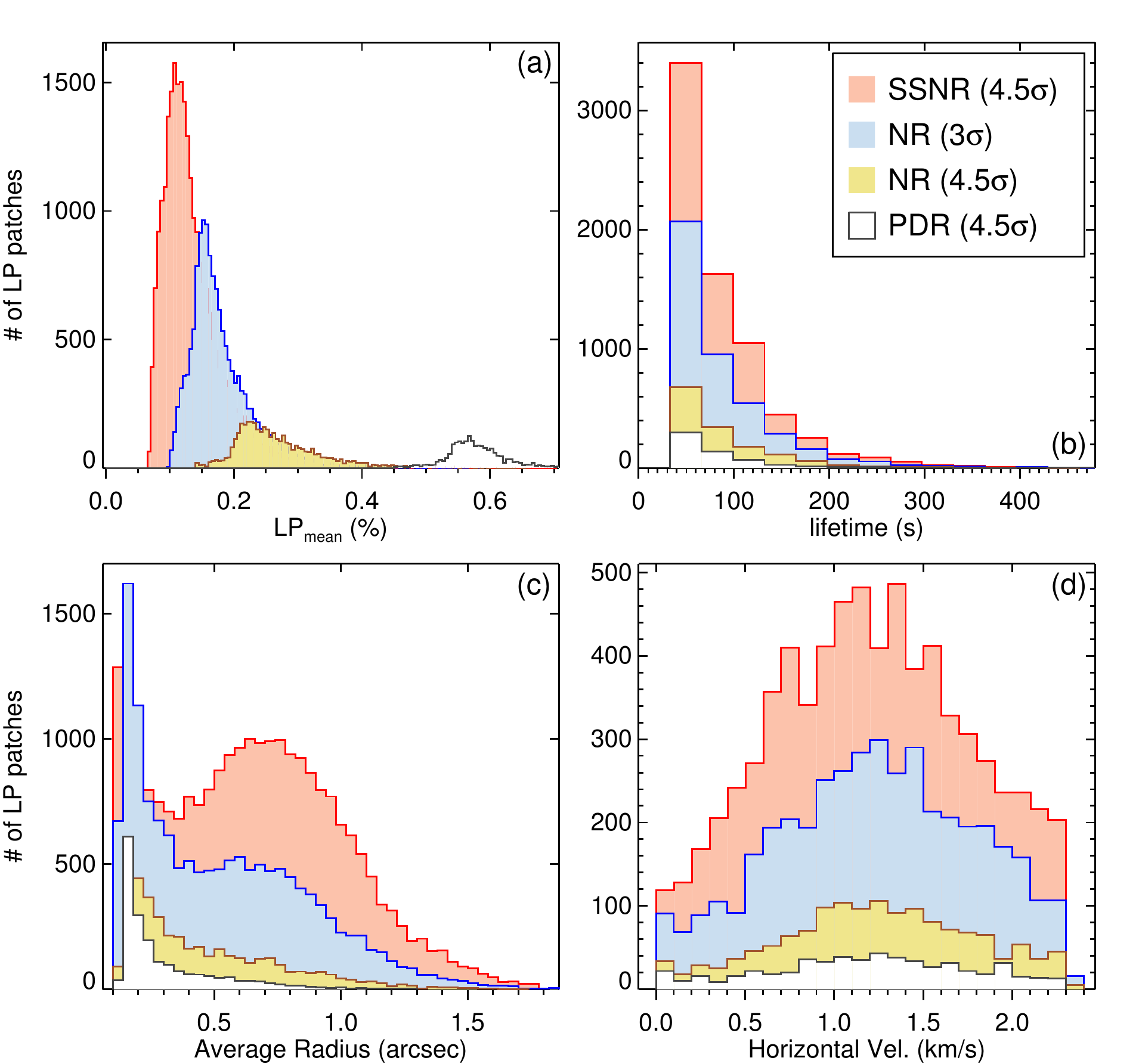}
              }
     \vspace{0.01\textwidth} 
	
	\caption{Distributions of mean LP signal (a), lifetime (b), average size (radius; (c)), and horizontal velocity (d) of the detected linear polarization features from differently treated data sets, \textit{i.e.}, PDR, NR and SSNR data with noise levels ($\sigma$) of $1.1\times 10^{-3}$, $3.4\times 10^{-4}$ and $1.7\times 10^{-4}$ in terms of continuum intensity, respectively (see main text).
        }
   \label{histogram}
\end{figure}

\begin{figure}
   \centering
	\includegraphics[width=\linewidth,clip=]{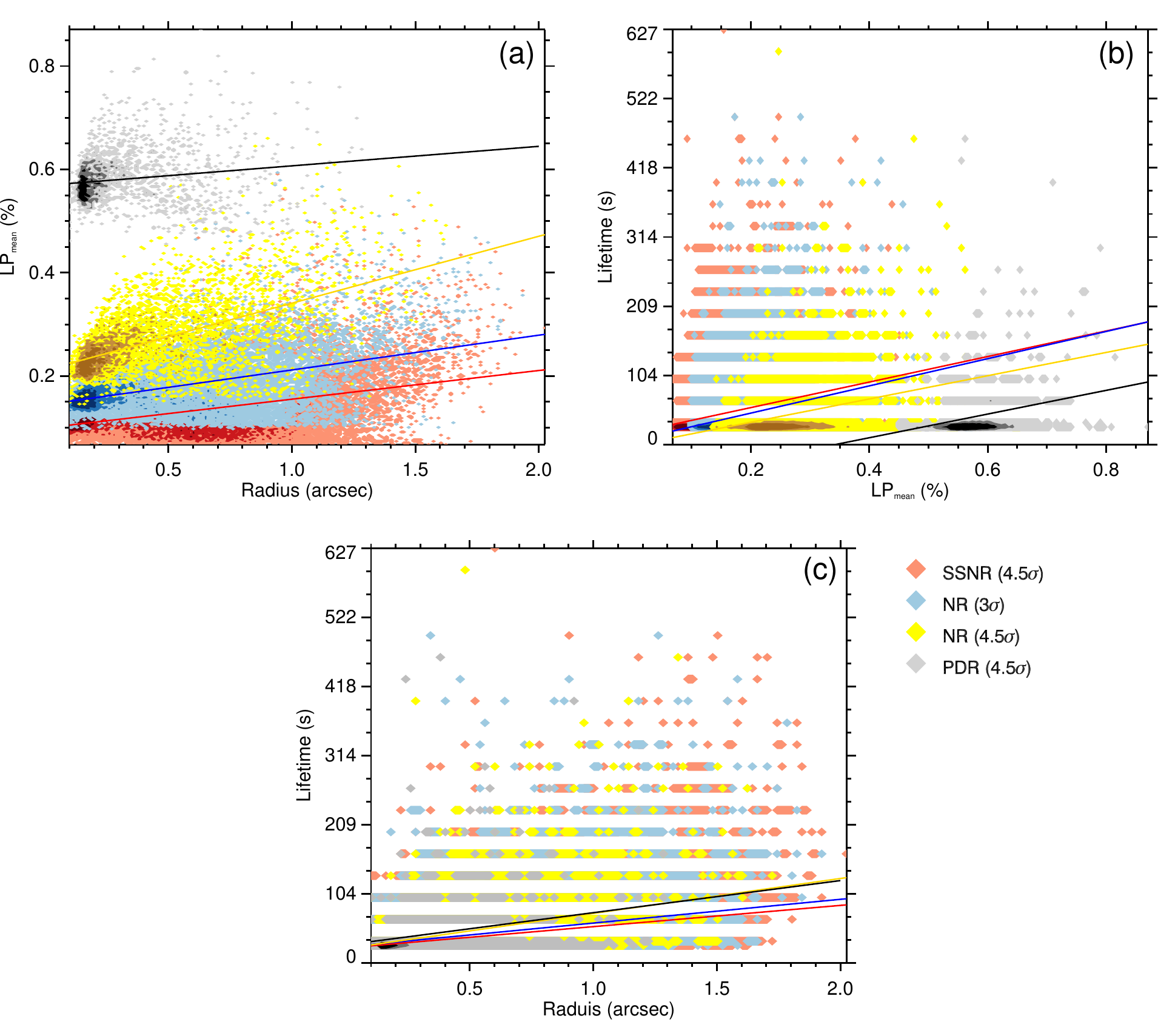} 
		\caption{Relationships between various pairs of parameters of the LPFs, detected in differently treated data sets (see main text).
		}
	\label{scatter}
\end{figure}

The LP signal, size, horizontal velocity, and lifetime of the detected LPFs from the differently treated data (\textit{i.e.}, see Table~\ref{table:LPFnumbers}) were determined during the detection and tracking procedures. The distributions of these parameters are shown in Figure~\ref{histogram}, and the relationships between pairs of the physical quantities are plotted in Figure~\ref{scatter}. Below, these parameters are individually discussed in detail.

\textit{\textbf{LP Signal}}. The LP values of all pixels within the area of an LPF is extracted from the LP maps, \textit{i.e.}, the pixel values of the detected features are individually stored. The mean and maximum LP values of each LPF are calculated for further statistics, as well as relationship studies between the various parameters. 
The histograms of mean LP signals (Figure~\ref{histogram}(a)) are all skewed to greater signal values. It is evident that by increasing the noise level, only features with stronger signals are detected. A comparison between the different histograms in Figure~\ref{histogram}(a) also shows the effect of S/N on the number of the identified LPFs. The distribution of the mean LP signals of the features found in the LP\textsubscript{SSNR} (\textit{i.e.}, the least noisy data set) ranges between 5$\sigma$\textsubscript{LP\textsubscript{SSNR}} and 11$\sigma$\textsubscript{LP\textsubscript{SSNR}}, peaking at $\approx8\sigma$\textsubscript{LP\textsubscript{SSNR}} (\textit{i.e.}, $1.3\times10^{-3}~I_{c}$).

\textit{\textbf{Lifetime}}. About $65\%$, in SSNR and NR data, and $75\%$, in PDR data, of the total detected features are observed in only one frame (\textit{i.e.}, they have a maximum lifetime of about 33~s). The lifetime of the rest of the other features are found from the tracking procedure, \textit{i.e.}, the time difference between the detected feature in the last frame and the first one. Figure~\ref{histogram}(b) shows the lifetime distribution of the LPFs which live longer than one frame. Thus the histograms are limited at 33~s in their lower levels.

These histograms indicate that lifetimes of the most of the LPFs that live longer than one frame range between 1 and 8~min. The lifetimes of the LPFs show no significant correlation with the S/N. The average lifetime in all the data sets is about 70~s. However the highest signal to noise ratio in SSNR data leads to the highest fraction ($35\%$) of long-lived features (\textit{i.e.}, features that live more than 8~min) among all tracked patches. This may suggest that the relatively weak LP signals at the beginning and/or at the end of the LPFs' life are being uncovered to a rather large extent when they are detected in images with larger S/N.

\textit{\textbf{Size}}. The size (\textit{i.e.}, the cross-section) of an LPF is defined as the number of pixels that are included in a feature boundary, defined based on a certain signal threshold. Figure~\ref{histogram}(c) shows the distribution of mean radius, computed from their areas (by assuming a circular shape). The LPFs in all the four data sets appear to have radius in the order of $0.1-1.5$~arcsec, comparable to the size of granules. It is also found that the higher S/N uncovers a larger (hidden) part of the LPFs, thus a larger number of features with bigger sizes are found in the SSNR data compared to the other (noisier) data sets. It is similarly the case when comparing the size distribution of LPF\textsubscript{NR} with S/N threshold of 3 and 4.5, the former includes many more larger features.

\begin{figure} 
   \centerline{\hspace*{0.0\textwidth} 
                          \includegraphics[width=1.\textwidth,clip=]{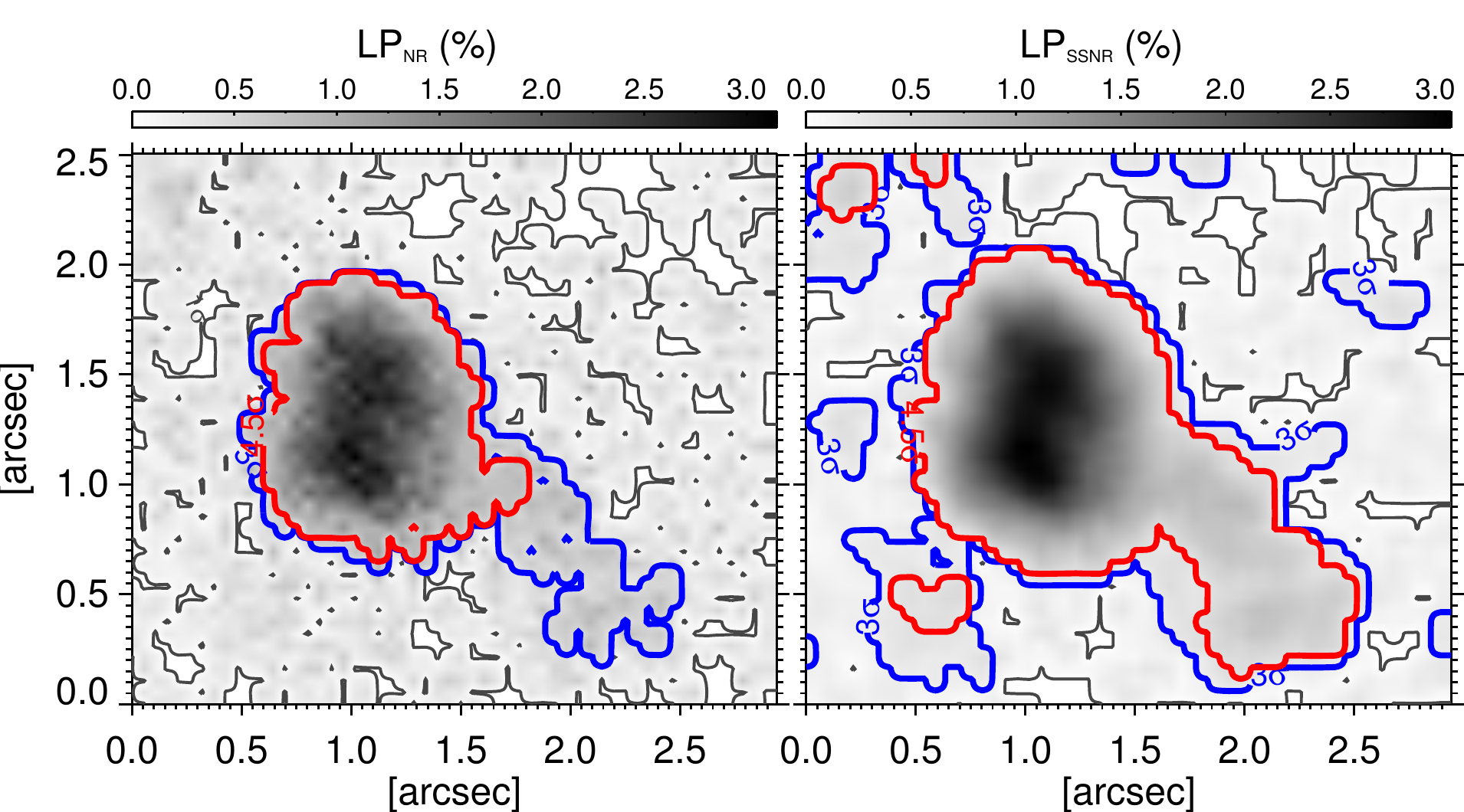}
              }
     \vspace{0.0001\textwidth} 
               
	\caption{A linear polarization feature (LPF) in non-reconstructed (NR; left) and spatially smoothed non-reconstructed (SSNR; right) images. The black, blue and red contours mark the corresponding $1~\sigma_{LP}$, $3~\sigma_{LP}$, and $4.5~\sigma_{LP}$ noise levels (see Table~\ref{table:LPFnumbers}). The differences between size and number of the detected features in the two panels are evident. 
        }
   \label{ind_feature}
\end{figure}

The larger size of features revealed by higher S/N (or lower signal threshold) may indicate that the LP signal radially reduces from the center of the LPFs toward the edges of the patches. Figure~\ref{ind_feature} illustrates the effect of these two factors on the size of a typical patch. The detected feature with the signal threshold of $4.5~\sigma_{LP}$ (red contour) is smaller in the NR map (left panel), compared to that found in the SSNR image (right panel). Also, a similar size difference is found for a lower signal threshold of $3~\sigma_{LP}$, shown with the blue contour. The LPF detected with the latter signal threshold encompasses a larger region than the case of applying $4.5~\sigma_{LP}$ threshold on the same map. In addition, the LP\textsubscript{SSNR} includes more than one isolated feature at both signal thresholds of $3~\sigma_{LP}$ and $4.5~\sigma_{LP}$ (\textit{i.e.}, the separated islands shown with the blue and red contours, respectively), whereas only one feature is identified in the LP\textsubscript{NR} image.

\textit{\textbf{Horizontal Velocity}}. The instantaneous horizontal velocity is determined for the tracked features (\textit{i.e.}, the features that lived for more than one frame). It is obtained by dividing the distance that each LPF moves between consecutive frames by their observing time-difference (\textit{i.e.}, the cadence of the observations; 33~s). 

We note that the tracking algorithm considers a feature to be the same as in the previous frame if it is spatially located in a small area around it. This condition has inevitably lead to an upper limit for the determined horizontal velocities. Although this criterion may exclude the very fast-moving features, it has secured our detection from mixing of the apparently close features. In addition, the horizontal velocity has been only measured for the LPFs whose lifetimes are longer than 33~s (\textit{i.e.}, when they are observed, at least, in two consecutive frames).

Figure~\ref{histogram}(d) shows the distribution of horizontal velocity of the detected LPFs from the differently treated data sets. The four histograms show a nearly normal distribution. They all range between $0-2.4$~km~s\textsuperscript{-1}, and peak at a nearly same mean velocity of 1.2~km~s\textsuperscript{-1}.\\

Figure~\ref{scatter} demonstrates the scatter density plots of the various parameters against each other. The points with darker colors show densely scattered areas. In Figure~\ref{scatter}(a) the distribution of mean LP of the tracked LPFs are plotted with respect to their mean radius, averaged over the course of their lifetimes. The linear fits to the data points show a direct correlation between the $LP_{mean}$ and mean radius of the LPFs in all the four differently treated data sets. 

A direct correlation also exists in case of $LP_{mean}$ versus lifetime as well as the average radius versus lifetime, respectively, shown in Figure~\ref{scatter}(b) and (c). The lifetime scatter plots in Figure~\ref{scatter} are binned to the cadence of the observations, \textit{i.e.}, 33~s.

These correlations indicate that the bigger features pose, on average, larger LP signals and tend to live longer on the solar surface.

No clear relationships between the horizontal velocity of the LPFs and their other properties was found.

\subsection{Stokes Inversion}
   \label{Sbb:inversion}

We also use the results of Stokes inversion (of the PDR data) to obtain additional physical properties of the detected LPFs as well. These include LOS velocity, magnetic field strength, field inclination, and temperature. We use the results of the SPINOR inversion code~\citep{Solanki1987,Frutiger2000,Berdyugina2003}. The code used the Harvard Smithsonian Reference Atmosphere (HSRA;~\citealt{Gingerich1971}) as the initial model atmosphere, and performed height-dependent computation of temperature at three nodes along the depth scale (at $log [\tau_{500~nm}]$=0, $-0.9$, and $-2.5$). The other parameters were calculated at one node. For details on the code and the specifications applied on the same data see \citet{Kahil2017}. It is noted that the inversion code have assumed a unity magnetic filling factor.

We extract the physical parameters from the corresponding pixels (of the center of gravity of LP signal) of the LPFs from the results of the Stokes inversion. Although these may not represent the physical properties of all pixels across individual LPFs, the parameters correspond to pixels with the largest S/N in each feature.

\begin{figure}
   \centerline{\hspace*{0.01\textwidth} 
                          \includegraphics[width=0.99\textwidth,clip=]{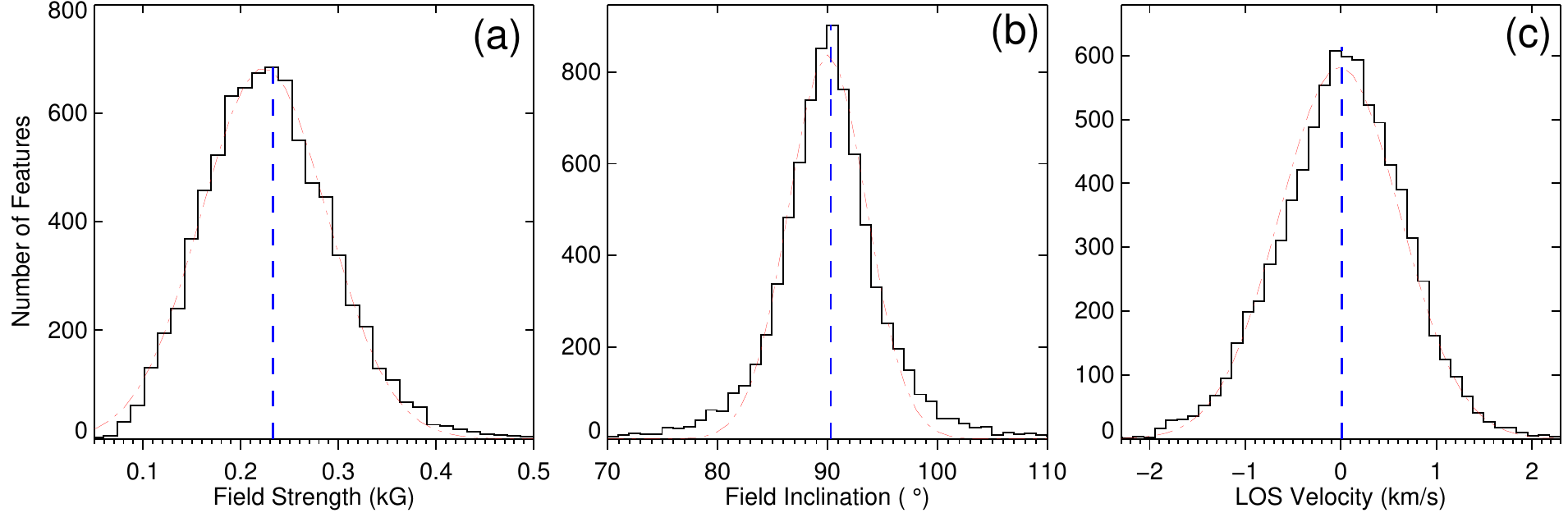}
              }
     \vspace{0.01\textwidth} 
	
	\caption{Distributions of the field strength (a), field inclination (b), and LOS velocity of the LPFs from the PDR Stokes inversion with the SPINOR code.
        }
   \label{inv_hist}
\end{figure}    

\subsubsection{Physical Properties from Stokes Inversion}
   \label{Sbb:F-char2}

The distributions of the magnetic field strength, field inclination, and the LOS velocity of the LPFs, from the SPINOR inversion code, are plotted in Figure~\ref{inv_hist}(a)-(c), respectively. They show that the magnetic field at the position of LPFs are almost horizontal, with a rather narrow histogram peaking at an inclination angle of 90 degrees (Figure~\ref{inv_hist}(a)). The LPFs are found to pose hG fields at their center of gravity (Figure~\ref{inv_hist}(b)). We note, however, that the field strength has been likely underestimated at individual pixels (they are spatially unresolved). 
The histogram of LOS velocity of the LPFs, shown in Figure~\ref{inv_hist}(c), demonstrates that they are located on both the center of the granules, associated to upflows, and granular edges, where downflows have been reported (\citealt{roudier11,Nordlund2009}). Their distributions over the two regions are nearly equal in the results of the SPINOR code.

\begin{figure}
   \centerline{\hspace*{0.01\textwidth} 
                          \includegraphics[width=1.0\textwidth,clip=]{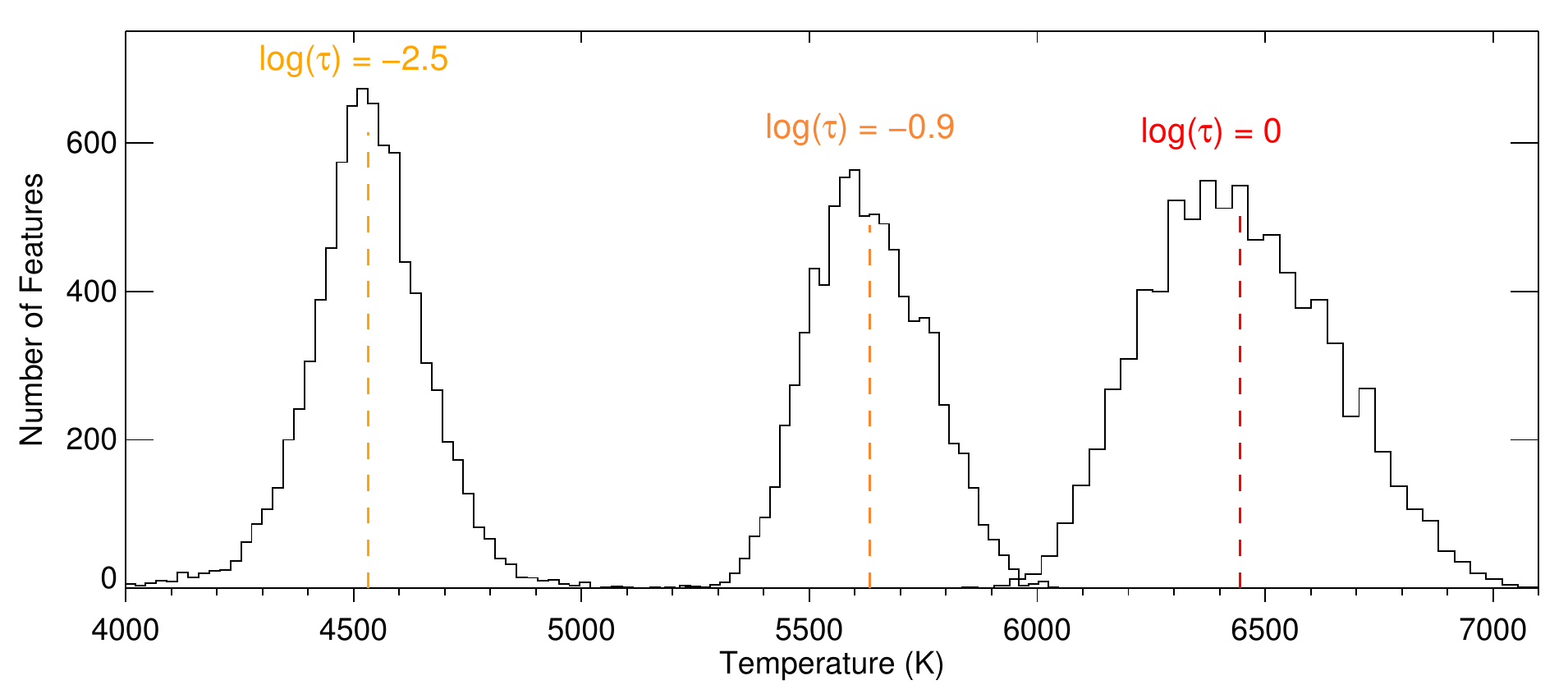}
              }
     \vspace{0.01\textwidth} 
	
	\caption{Distributions of temperature at the location of the LPFs, from the results of the SPINOR inversion code. The distributions at three optical depths (labeled above the peak of each histogram) are plotted.
        }
   \label{inv_hist2}
\end{figure}

The distributions of temperature at the position of the detected LPFs are plotted at three optical depths in Figure~\ref{inv_hist2}. We preciously found that the LPFs were, on average, brighter than the quiet Sun (see Section~\ref{Sb:F-intensity}). This implies that the LPFs are likely located over the solar granules. This agrees with the decrease of their temperatures, on average, with height (see Figure~\ref{inv_hist2}), that is expected from the standard model atmosphere, such as FALC \citep{Fontenla2006}.

\begin{figure} 
   \centerline{\hspace*{0.0\textwidth} 
                          \includegraphics[width=1.0\textwidth,clip=]{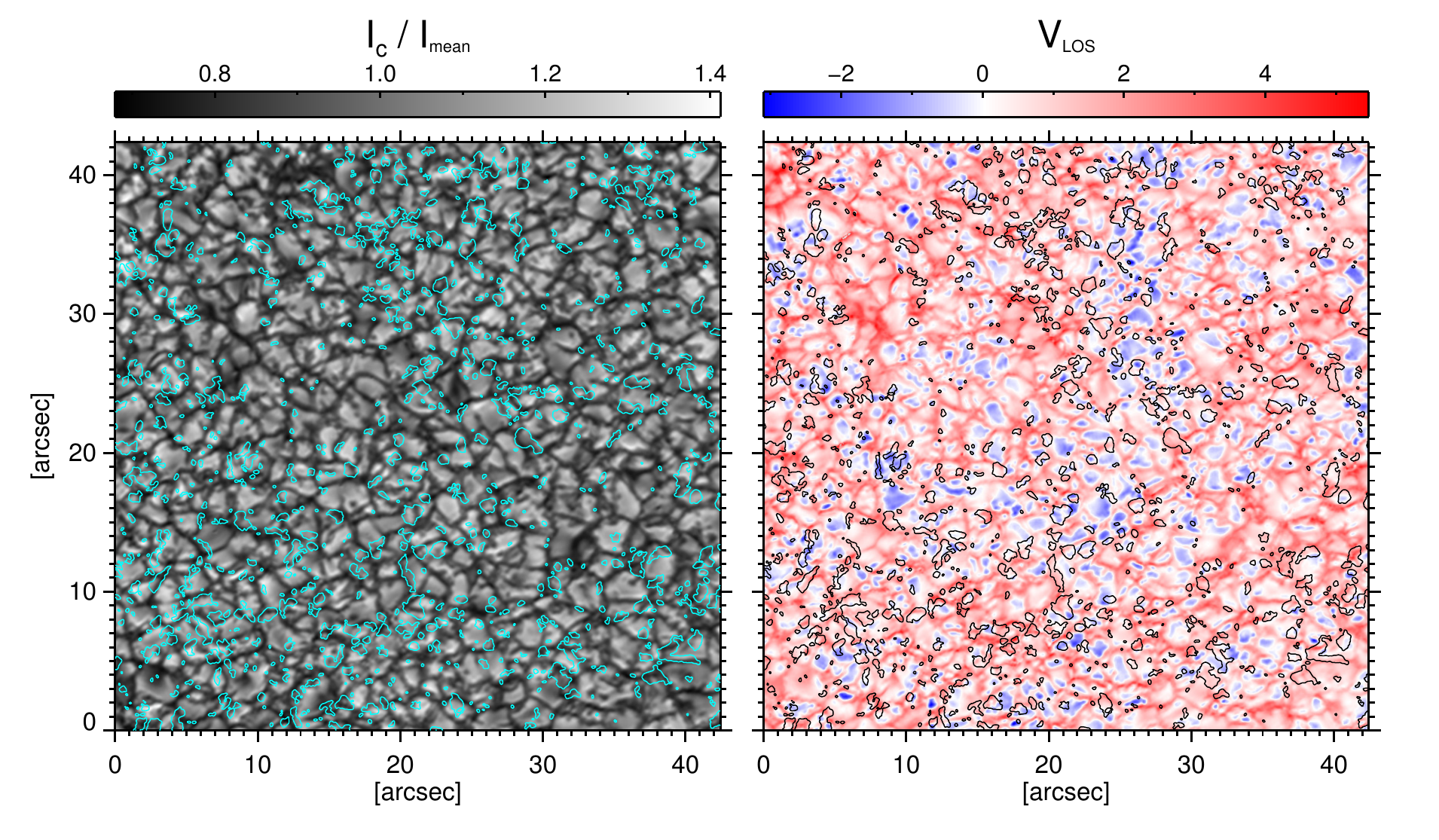}
              }
     \vspace{0.01\textwidth} 
	\caption{Spatial locations of the linear polarization features (LPFs) on a continuum intensity image (\textit{left}), and on its corresponding LOS velocity map (\textit{right}) of the first frame of PDR data set.
        } 
   \label{IcVlos}
\end{figure}

To also visually inspect the locations of the LPFs in respect to the granulation pattern, we have plotted the identified LPFs on a continuum intensity image as well as its corresponding LOS velocity map in the left and right panels of Figure~\ref{IcVlos}, respectively. These show that the LPFs are mostly located over the granules. A small fraction of some of the LPFs seems to be extended to the intergranular areas.

\section{Conclusions}
	\label{conculsion}

We have carried out a thorough study of (statistical properties of) linear polarization features (LPFs) by exploiting the high spatial and temporal-resolution observations of the quiet-Sun internetwork, obtained with {\sc Sunrise}/IMaX. The LPFs are defined as areas with at least 10 contiguous pixels whose LP signals are above a particular threshold level. We inspected the effect of the S/N in our study by analyzing the LPFs in three differently treated data sets (with different noise levels; see Table~\ref{table:noiselevels}) and applying various signal thresholds.

We found a total number of $\approx25100$ individual LPFs (when each LPF was only counted once during the course of its lifetime) during the 31~min time-series of images of the SSNR data (\textit{i.e.}, consisting 58 frames of $45''\times45''$). This is, to our knowledge, the largest number of individual LPFs found so far, with the rate of occurrence on the order of $8\times10^{-3}$~s$^{-1}$\,arcsec$^{-2}$. This rate of occurence is larger, by an order of magnitude, than that of found by \cite{danilovic10}. We should, however, note that the rate of occurence strongly depends on the definition of the features under study. The detected LPFs from the SSNR images found to have an average radii on the order of $0.5-1$~arcsec and cover about 10.3\% of quiet-Sun internetwork.

Spatially smoothing of the NR data, resulting in a higher S/N, revealed detection of larger LPFs. The phase-diversity reconstructed data, leading to a higher spatial resolution compared to the NR images, resulted in identification of very small LPFs with an average diameter of $\approx0.2$~arcsec. The S/N were also found to influence the distribution of lifetime of the LPFs, so longer-lived features were detected in data sets with higher S/N. The lifetime distributions of the LPFs (peaking at about 1~min) drop exponentially for all the differently treated data sets. The short lifetimes of our LPFs are in agreement with the transitory nature of these features reported by \cite{lites96} and \cite{danilovic10}, but are smaller than the average lifetime obtained by \cite{pontieu02}, \cite{ishikawa08b} and \cite{Jin09}. The lifetimes of our LPFs are found to be correlated with their both mean LP signals and mean sizes (when averaged over their lifetimes).

The LPFs in our study horizontally move with an average speed of 1.2~km~s\textsuperscript{-1}, with nearly normal distributions ranging between $0-2.5$~km~s\textsuperscript{-1}. We found no correlation between the S/N of the employed data sets and the distributions of horizontal velocity of the LPFs.

We also examined the effect of signal threshold on the NR data. With a threshold of 4.5$\sigma_{LP}$, we found the smallest number of detected LPFs per frame (535 features on average) which occupies only $1.1\%$ of the total area. These LPFs emerge with a size of $\approx0.6$~arcsec and a minimum LP signal of about $2.7\times10^{-3}~I_{c}$. They reach a peak size and LP SIGNAL of $0.7$~arcsec and $2.9\times10^{-3}~I_{c}$ during their lifetimes, respectively, and disappear with almost having the same sizes and LP signals as the moment of their appearance on the solar surface. A total number of 4092 individual LPFs were tracked in the NR time-series of images, among which, about $25-35\%$ lived longer than 33~s (\textit{i.e.}, they were tracked in more than one frame). Applying a lower signal threshold of 3$\sigma_{LP}$ lead to a greater number of detected features in each frame, compared to those found with the signal threshold of $4.5\sigma_{LP}$. These LPFs (in NR images) cover about $5\%$ of the area, with the percentage of the number of features living longer than 33~s equal to $29\%$. Due to the lower signal threshold, the weaker features were also detected. These features have smaller mean LP signal than that of found for the LPFs with a signal threshold of $4.5\sigma_{LP}$, while the sizes of the former are, on average, bigger than the latter. This indicates that the LP signal decreases from the center of the patches toward their boundaries, which can, however, be a result of the limited spatial resolution of the data set.

We also inspected the effect of the spatial resolution by investigating the LPFs in the PDR maps (\textit{i.e.}, with a larger spatial resolution, by a factor of two, compared to the NR images), which also has the largest noise level in our samples due to the reconstruction process (which amplifies both the signal and noise at the same time; see Table~\ref{table:noiselevels}). The LPFs studied in this data set (with a signal threshold of 4.5$\sigma_{LP}$) have an average LP signal of $\approx3.9\times10^{-3}~I_{c}$. They were found to be, on average, the shortest lived and the smallest features among the LPFs detected in all the other data sets.

We found a weak but positive correlation (due to the large scatter shown in Figure~\ref{scatter}) between the LP signal, size and lifetime of the features in all SSNR, NR, and PDR images. The \textit{Pearson correlation coefficient}~($r$) for the scatter density plots shown in Figure~\ref{scatter}, has been calculated for the most populated sample (\textit{i.e.}, the SSNR LPFs) as

\begin{equation}
\label{eq:coef}
r=\frac{\sum_{i=0}^{n} (x_i-\bar{x})(y_i-\bar{y})}{\sqrt{\sum_{i=0}^{n} (x_i-\bar{x})^2}{\sqrt{\sum_{i=0}^{n} (y_i-\bar{y})^2}}}
\end{equation}

\noindent where, $x$ and $y$ represent the properties scattered in each panel of Figure~\ref{scatter} and $n$ is the number of individual tracked LPFs in SSNR data. The coefficient values for the scatter plots of LP$_{mean}$-Radius (Figure~\ref{scatter}(a)), Lifetime-LP$_{mean}$ (Figure~\ref{scatter}(b)) and lifetime-Radius (Figure~\ref{scatter}(c)) are +0.41, +0.59 and +0.37, respectively. In other words, the LPFs with larger LP signal tend to grow to larger sizes and live longer, whereas the relatively small features with relatively weak LP signal are more transient. The horizontal velocity with which features move along the surface, shows no correlation with the other parameters in either of the differently treated data sets.

Furthermore, from the results of a Stokes inversion of the PDR data, we found that the LPFs have a magnetic field strength in the range of 50-500~G, peaking at 230~G. This agrees with the hG horizontal magnetic fields reported by \cite{lites96} and \cite{meunier98}. The mainly horizontal orientation of LPFs in the photospheric IN regions (\citealt{livingston71,livingston75,martin88,Lin95,Lin99,Lites2017}) was observed as the predominant inclination of 90$^{\circ}$, in a range of 70 to 110 degrees.  On average, the temperature at the location of the LPFs decreases with height in the solar photosphere.

Interactions between the LPFs, such as merging and splitting, as well as variations of the properties of the LPFs with time were not investigated here and are the subject of a future study.

\begin{acks}
We are grateful to A. Lagg for detailed and helpful comments on the manuscript. S.K. acknowledges the support from CHROMATIC project (2016.0019) funded by the Knut and Alice Wallenberg foundation. S.J. acknowledges support from the European Research Council (ERC) under the European Union’s Horizon 2020 research and innovation program (grant agreement No. 682462) and from the Research Council of Norway through its Centres of Excellence scheme, project number 262622. M.T.M. acknowledges support from the Iranian National Research Institute for Science Policy, Ministry of Higher Education. The German contribution to {\sc Sunrise} is funded by the Bundesministerium f\"{u}r Wirtschaft und Technologie through Deutsches Zentrum f\"{u}r Luft- und Raumfahrt e.V. (DLR), grant No. 50 OU 0401, and by the Innovationsfond of the President of the Max Planck Society (MPG). The Spanish contribution has been funded by the Spanish MICINN under projects ESP2006-13030-C06 and AYA2009-14105-C06 (including European FEDER funds). The HAO contribution was partly funded through NASA grant NNX08AH38G. 
\end{acks}



\bibliographystyle{spr-mp-sola}
\bibliography{Linear_Polarisation_Features}  
    
\end{article} 

\end{document}